\documentclass[prd,aps,showpacs,showkeys,twocolumn,10pt,nofootinbib]{revtex4}
\usepackage{amssymb,amsmath,amsthm}
\usepackage{mathrsfs}
\usepackage[applemac]{inputenc}

\usepackage{graphicx}
\usepackage{dcolumn}
\usepackage{bm}

\usepackage{xcolor}

\begin{document} 
\preprint{IPARCOS-UCM-23-012}

\title{Graviton-photon oscillation in a cosmic background for a general theory of gravity}

 \author{Jos\'e A. R. Cembranos} 
\email{cembra@ucm.es}
\author{Miguel Gonz\'alez Ortiz}
\author{Prado Mart\'in-Moruno}
\email{pradomm@ucm.es}
\affiliation{Departamento de F\'isica Te\'orica and IPARCOS,
Universidad Complutense de Madrid, E-28040 Madrid, Spain}

\date{\today}

\begin{abstract}
Graviton-photon oscillation is the conversion of gravitational waves to electromagnetic waves and vice versa in the presence of a background electromagnetic field.
We investigate this phenomenon in a cosmological scenario considering a background cosmic magnetic field and assuming different gravitational frameworks. 
We obtain the damping term that characterizes the attenuation of the conversion probability in cosmological backgrounds. This is a general feature that is present even for standard General Relativity. Furthermore, we show that the effects of decoherence, which are due to the interaction with the cosmological expansion and with the additional degrees of freedom of alternative theories of gravity, can be relevant to the phenomenon of graviton-photon mixing.
\end{abstract}


\maketitle

\section{Introduction}

Observational data based on electromagnetic waves (EMWs) have unveiled countless secrets of the cosmos. They have provided us with information about the history of the Universe and the dynamics of astrophysical objects. This knowledge has allowed us to test General Relativity (GR), concluding that it can suitably describe our cosmos assuming the existence of two dark cosmic components. In 2016 the observations came from a different domain for the very first time and GR passed one of its latest checks due to the detection of gravitational waves (GWs) \cite{LIGOScientific:2016aoc}. This detection was just the beginning of the era of gravitational wave measurements opening a new channel for multi-messenger astronomy.

Despite the great success of GR there are some issues that deserve further attention. On the one hand, at large scales it is necessary to assume the existence of dark matter and dark energy to describe the observational data.
As it is well known, most of the matter of the Universe has to be dark, but the nature of the dark matter particle remains elusive. 
Furthermore, dark energy should be now almost the 70\% of the cosmic energetic content to suitably drive the current accelerated cosmic expansion. However, it is not clear that the final description of dark energy is just a cosmological constant.
On the other hand, in the microscopic world of particles and high energies GR is thought to be inadequate. Any attempts to quantize and unify gravity with the Standard Model has been unsuccessful.
So, in recent years alternative theories of gravity (ATGs) \cite{Clifton:2011jh,EuclidTheoryWorkingGroup:2012gxx,CANTATA:2021ktz} have been explored in great detail as possible candidates to describe the gravitational phenomena at all scales.

A popular group of ATGs is formed by scalar-tensor theories of gravity, which introduce a non-minimally coupled scalar field into the action. In this framework Horndeski theory is the most general family of models leading to second-order differential equations \cite{Horndeski,Deffayet:2011gz,Kase:2018aps}. On the other hand, bimetric gravity theories introduce a second metric tensor in the formulation of the theory. These theories are related to massive gravity, which describes a single massive spin-2 field \cite{deRham:2014zqa}.  Other well-studied theories are $f(R)$-theories of gravity \cite{Nojiri:2006gh,Sotiriou:2008rp}. These theories introduce higher-order terms into the gravitational action with a generic function of the Ricci scalar. (We do not intend to provide an exhaustive list of ATGs in this introduction. The interested reader can check, for example, reference  \cite{CANTATA:2021ktz}.)

There is an interesting theoretical prediction of GR regarding the interaction of GWs and EMWs that has long been known: graviton-photon conversion in an electromagnetic background \cite{Gertsenshtein,Lupanov}. 
This is a phenomenon of conversion between GWs and EMWs that is produced when the waves cross an external electromagnetic field. (A non-evolving electromagnetic background is usually assumed, simplifying the mathematical treatment.)
In addition, when the waves propagate through a (real or effective) dielectric medium, there is no complete conversion, but graviton-photon oscillation or mixing \cite{Zeldovich,Raffelt}. This effect is analogous to neutrino oscillations, for example. This is, the evolution of the system at first order in perturbations is such that gravitational and electromagnetic degrees of freedom are mixed as a single superposed state throughout its propagation. From a cosmological point of view it is of special interest the phenomenon that can take place due to the propagation of primordial GWs through the cosmic magnetic field \cite{Dolgov:2012be,Dolgov:2013pwa} (see also references \cite{Ejlli:2018hke,Ejlli:2020fpt} for more recent studies). The hypothetical detection of EMWs generated in this process would allow us to measure indirectly GWs that are beyond the scope of current detectors. 
Moreover, if those waves were not observed in the range that they are predicted, it may indicate the absence of the corresponding primordial GWs or point towards some modification of the gravitational theory. In this respect, the predictions about the efficiency of the conversion process in GR indicate that the sensibility of the instruments could be too low to measure the effect of laboratory experiments \cite{Boccaletti}, although it could left footprints in the radiation cosmic background (excluding their relevance in the microwave) \cite{Dolgov:2012be}.

We want to stress that this phenomenon is predicted in GR even in regions where the refractive index equals unity, in which case there is just conversion between EMWs and GWs. This is the case when we just consider classical electrodynamics and GR. Therefore, in order to get a mixture in GR and not just conversion, it is necessary to take into account terms of greater perturbative order into the Maxwell Lagrangian, such as the Euler-Heisenberg Lagrangian \cite{Raffelt}. This term describes nonlinear corrections and induces an effective refractive index for EMWs \cite{Dolgov:2012be,Dolgov:2013pwa}.

On the other hand, following the spirit of reference \cite{Cembranos:2018jxo}, one can also consider that GWs propagate with an effective refractive index. This situation does not appear in GR for vacuum solutions, since GWs propagate in vacuum at the speed of light. When they propagate in a FLRW spacetime, however, the expansion of the Universe generates attenuation in the amplitude of the wave \cite{Caprini:2018mtu}. Considering that the gravitational phenomena are described by a given ATG, the situation can be much more varied. Propagation of GWs in ATGs can include, for example, a mass term, an attenuation term, or variations in the speed of propagation \cite{Saltas:2014dha,Bettoni:2016mij}. 
By analogy with electromagnetism, where these terms arise when EMWs propagate through a dielectric media, one can think in a medium in which GWs propagate formed by the additional degrees of freedom of the ATGs with respect to GR. These effective media are called diagravitational media \cite{Cembranos:2018lcs}. Multi-messenger astronomy \cite{LIGOScientific:2017vwq} has already allowed to conclude that GWs propagate today at the speed of light, ruling out a large number of ATGs as dark energy mimickers \cite{Lombriser:2016yzn,Ezquiaga:2017ekz,Creminelli:2017sry,Sakstein:2017xjx}. However, other theories are still compatible with this constraint. 
For instance, theories of massive gravity induce a dispersive medium \cite{deRham:2014zqa}. An attenuation term may also appear in the refractive index, leading to wave amplification or absorption.  This term, which is typically linked to variations in some fundamental quantity of the ATG \cite{Barrow:1993ad}, combines with the cosmic attenuation in cosmological backgrounds.
In addition, Lorentz violating ATGs entail modified dispersion relations
\cite{Mirshekari:2011yq}. All these different terms can be included in a refractive index for the diagravitational medium  \cite{Cembranos:2018lcs}. Therefore, the predictions for the phenomenon of graviton-photon oscillations may be affected in ATGs due to the existence of this effective medium.

Graviton-photon mixing in ATGs was first considered in reference \cite{Cembranos:2018jxo}, paying particular attention to the case of a real refractive index when obtaining the conversion probability. In this work we shall investigate a more general framework in detail, taking into account in discussing the effects of a cosmic background.
This work can be outline as follows: In section \ref{secII} we present the results that we have obtained for the phenomenon of graviton-photon mixing following a procedure based in studying the mixing probability. 
In section \ref{secIIA} we focus on GR and obtain the mixing matrix of the graviton-photon system in a cosmological scenario, extending the results of reference \cite{Dolgov:2012be} to propagation in a FLRW spacetime. 
In section \ref{secIIB} we discuss how the system is modified when considering theories beyond GR, as in reference \cite{Cembranos:2018jxo}, but now, for a cosmological background. 
In section \ref{secIIC} we obtain the mixing probability of primordial GWs and EMWs and estimate the effect in some ATGs. 
On the other hand, in section \ref{secIII}, we consider the density matrix formalism taking into account effects of photons decoherence. We consider for the very first time decoherence effects affecting gravitational radiation and quantify the modification of the predictions of graviton-photon mixing for some ATGs. Finally, in section \ref{secIV}, we summarize our results.


\section{Graviton-photon mixing in a cosmological background}\label{secII}
In order to describe graviton-photon mixing we need to consider an action composed by two parts.
The gravitational part is the Einstein--Hilbert action if we assume GR, but has a different form for ATGs.
We consider that the action of the electromagnetic part is given by the Maxwell action plus the Euler--Heisenberg term. This is
\begin{equation}\label{eq2}
\begin{split}
    S_{em}=&-\dfrac{1}{4}\int d^4x \sqrt{-g}F_{\mu\nu}F^{\mu\nu}\\
    &+\dfrac{\alpha^2}{90m_e^4} \int d^4x \sqrt{-g}[(F_{\mu\nu}F^{\mu\nu})^2+\dfrac{7}{4}(\widetilde{F}_{\mu\nu}F^{\mu\nu})^2],
    \end{split}
\end{equation}
where we have introduced the usual electromagnetic field tensor $F_{\mu\nu}=\partial_{\mu}A_{\nu}-\partial_{\nu}A_{\mu}$.
Here, $\alpha=e^2/4\pi$ is the fine structure constant, $m_e$ is the electron mass and $\widetilde{F}_{\mu\nu}=\dfrac{1}{2}\epsilon_{\mu\nu\rho\sigma}F^{\rho\sigma}$ is the dual of $F_{\mu\nu}.$ The Euler--Heisenberg second order term describes quantum corrections to the classical electromagnetic action and it is only valid at the low frequencies regime, $\omega\ll m_e$. As we will consider frequencies up to $O(10^5)$eV for the GWs, we can assume that this approximation is valid.

On the other hand, we consider a perturbed flat FLRW spacetime. This can be expressed with
\begin{equation}\label{eq1}
    ds^2=a(\eta)^2[-d\eta^2+(\delta_{ij}+\kappa h_{ij})dx^idx^j],
\end{equation}
where we have introduced $\kappa=\sqrt{16\pi G}$ in order to have a dimensionless metric perturbation $\kappa h_{ij}$. 
This metric is expressed in the conformal time coordinate, $\eta$, related with the cosmological time as $dt=a(\eta)\,d\eta$. Moreover, we will make use of the Trace-Transverse gauge (TT), that is, $h_{0\mu}=\partial_{i}h_{ij}=h_i{}^i=0$; so only the spatial components are non trivial. Regarding the electromagnetic field, we assume that the electromagnetic tensor is composed by a background external contribution plus a variable small perturbation, $F_{\mu\nu}=F^{(e)}_{\mu\nu}+f_{\mu\nu}$. Thus, the term $f_{\mu\nu}$ suitably describes the photon field when $|f_{\mu\nu}|\ll |F^{(e)}_{\mu\nu}|$.

In order to get the equations of motion we expand metric (\ref{eq1}) in the action (\ref{eq2}) up to second order in $f_{\mu\nu}$, keeping also the mixing second order in $h_{\mu\nu}$ and $f_{\mu\nu}$ and neglecting the complete gravitational second order term in $h_{\mu\nu}$. As we are interested only in terms that describe the evolution of perturbations of the graviton and photon fields, we ignore the terms of first order in the action, which will lead to the Einstein equation of the background. The electromagnetic action including only these second order terms in perturbations is
\begin{equation}\label{eq3}
\begin{split}
    S_{em}^{(1)}&=-\dfrac{1}{4}\int d^4x f_{\mu\nu}f_{\rho\sigma}\eta^{\mu\rho}\eta^{\nu\sigma}+\dfrac{\kappa}{2}\int d^4xh_{\mu\nu}T^{em}_{\rho\sigma}\eta^{\mu\rho}\eta^{\nu\sigma}
    \\&
    +\dfrac{\alpha^2}{180m_e^4} \int d^4x \frac{1}{a^4}
    [4(F^{(e)}_{\mu\nu}f_{\rho\sigma}\eta^{\mu\rho}\eta^{\nu\sigma})^2
    \\&
    +4(F^{(e)}_{\mu\nu}F^{(e)}_{\rho\sigma}\eta^{\mu\rho}\eta^{\nu\sigma})
    (f_{\alpha\beta}f_{\gamma\delta}\eta^{\alpha\gamma}\eta^{\beta\delta})
    \\&
    +7(\widetilde{F}^{(e)}_{\mu\nu}F^{(e)}_{\rho\sigma}\eta^{\mu\rho}\eta^{\nu\sigma})
    (\widetilde{f}_{\alpha\beta}f_{\gamma\delta}\eta^{\alpha\gamma}\eta^{\beta\delta})
    \\&
    +7(\widetilde{F}^{(e)}_{\mu\nu}f_{\rho\sigma}\eta^{\mu\rho}\eta^{\nu\sigma})^2
    ],
    \end{split}
\end{equation}
where the electromagnetic energy-momentum tensor needs to be introduced at first order in the electromagnetic perturbation:
\begin{equation}\label{eqEMT}
\begin{split}
    T^{\rm em}_{\mu\nu}
    &=F^{(e)}_{\mu\rho}f_{\sigma\nu}\eta^{\rho\sigma}
    +f_{\mu\rho}F^{(e)}_{\sigma\nu}\eta^{\rho\sigma}
    -\frac{1}{2}\eta_{\mu\nu}F_{\alpha\gamma}f_{\beta\delta}\eta^{\alpha\beta}\eta^{\gamma\delta}.
    \end{split}
\end{equation}
Here, we have consistently neglected the contribution of the Euler--Heisenberg term.
In Eq. \eqref{eq3}, the dual tensors are defined with respect to the Minkowski metric $\eta_{\mu\nu}$.
Working with conformal time, the only dependence with the expanding geometry is reduced to the scale factor term in front of the Euler--Heisenberg Lagrangian. In fact, as it is well-known, due to the conformal invariance of the Maxwell action, there is not friction term associated to electromagnetic waves propagating in FLRW geometries \cite{Durrer:2013pga}. In the case of electromagnetic background fields, the Euler--Heisenberg contribution can introduce such a term. However, it will be proportional to the electromagnetic external field multiplied by the small parameter $\alpha^2$, and we will neglect it along the analysis of this work. In addition, we will assume that these external fields will be static (or slowly varying).

\subsection{Equations of motion in GR}\label{secIIA}
Let us consider that the gravitational phenomena are described by GR. Thus, the action for the metric field is given by the Einstein--Hilbert action
\begin{equation}\label{eq4}
    S_g=\dfrac{1}{\kappa^2}\int d^4x \sqrt{-g}R,
\end{equation}
where $R$ is the curvature scalar and $g$ is the determinant of the metric $g={\rm det}(g_{\mu\nu})$. The Riemann tensor is defined as $R^{\lambda}_{\mu\nu\sigma}=\partial_{\nu}\Gamma^{\lambda}_{\mu\sigma}-\partial_{\sigma}\Gamma^{\lambda}_{\mu\nu}+\Gamma^{\lambda}_{\alpha\nu}\Gamma^{\alpha}_{\mu\sigma}-\Gamma^{\lambda}_{\alpha\sigma}\Gamma^{\alpha}_{\mu\nu}$. 
Varying this action with respect to $h_{\mu\nu}$ and $A_{\mu}$ (where, from now on, $A_{\mu}$ denotes just the photon potential, which is associated to $f_{\mu\nu}$), one gets the coupled equations of motion for the graviton-photon system. 

The variation of the action with respect to $h_{\mu\nu}$ leads to \cite{Caprini:2018mtu}
\begin{equation}\label{5p}
\begin{split}
    a^2\Box \bar{h}_{\mu\nu}+2g^{\alpha\beta}\partial_{\alpha}a^2\nabla_{\beta}\bar{h}_{\mu\nu}&\\
    +\partial_{\alpha}\partial^{\alpha}a^2\bar{h}_{\mu\nu}-2a^2R^{\lambda}_{\mu\nu}{}^{\sigma}\bar{h}_{\lambda\sigma}&=-\kappa T_{\mu\nu}^{(1)},
\end{split}
\end{equation}
which is the Einstein equation with the electromagnetic energy-momentum tensor up to first order in the metric perturbation, and in the photon perturbation as the source. 
Note that in Eq. \eqref{5p} and from now on, we use the Minkowski metric $\eta_{\mu\nu}$, 
to raise, lower, and contract indexes.
We have defined the trace-reversed metric perturbation, $\bar{h}_{\mu\nu}= h_{\mu\nu}-\frac{1}{2}\eta_{\mu\nu}h$, which will be the new field representing the graviton. 
From now on, we will omit the bar. 
Since on the TT gauge only the spatial components are non-vanishing, equation (\ref{5p}) can be written as
\begin{equation}\label{eq6}
    h_{ij}^{''}-\nabla^2h_{ij}+2\mathcal{H}h_{ij}^{'}=\kappa T_{ij}^{(1)},
\end{equation}
where $\mathcal{H}=\dfrac{a'}{a}$ is the conformal Hubble rate and the prime denotes derivatives with respect to the conformal time. This is a wave equation with a dumping term that is proportional to the Hubble expansion rate. 
On the right hand side of (\ref{eq6}) we have the first order electromagnetic energy-momentum tensor as the source, whereas its background components are associated with the background Einstein equations. 
Now we assume that the external electromagnetic field in the chosen coordinates has only a magnetic component. Therefore, the first order energy-momentum tensor can be written as $T_{ij}^{(1)}=B_{(i}^eB_{j)}$, where the parentheses indicate symmetrization and $(e)$ denotes the component of the external magnetic field.

On the other hand, the equation for the propagation of the photon field is obtained by varying action (\ref{eq3}) with respect to $ A_{\mu} $. So, one obtains
\begin{equation}\label{eq7}
\begin{split}
  &  \nabla_{\mu}\left( f^{\mu\nu}-\dfrac{4\alpha^2}{90 a^4 m_e^4}[4F_{\alpha\beta}^{(e)}f^{\alpha\beta}F^{(e)\mu\nu}+7\widetilde{F}_{\alpha\beta}^{(e)}   f^{\alpha\beta}\widetilde{F}^{(e)\mu\nu}
  \right.
  \\&
  \left.
  4F_{\alpha\beta}^{(e)}F^{(e)\alpha\beta}f^{\mu\nu}+7\widetilde{F}_{\alpha\beta}^{(e)} F^{(e)\alpha\beta}\widetilde{f}^{(e)\mu\nu}
  ]\right) 
    =    \\
&\kappa\nabla_{\mu}\left( h^{\mu\beta}F^{(e)\nu}_{\beta}- h^{\nu\beta}F^{(e)\mu}_{\beta}\right).
    \end{split}
\end{equation}

In the first place, following the procedure commonly employed in the literature \cite{Dolgov:2012be,Dolgov:2013pwa,Ejlli:2018hke}, we use the gauge condition $\partial_iA^i=0$ and choose $A^0=0$. So, we can re-write the first term on the left hand side of equation (\ref{eq7}) in terms of the photon electromagnetic vector as $\Box A^j$ up to first order in perturbations. It has to be emphasized that this term does not lead to an absorption term due to the conformal symmetry of the field (as it can be explicitly checked using the tools summarized, for example, in reference  \cite{Durrer:2013pga}). In the second place,
 the right hand side of equation (\ref{eq7})  is a source term. This term contains the metric perturbations and the constant external field $F_{\mu\nu}^{(e)}$ (which only has spatial componentes). In the third place, the Euler--Heisenberg Lagrangian has led to the second term on the left hand side of equation (\ref{eq7}). 
As discussed in reference \cite{Brezin:1971nd} for the flat case, expanding the first part of this term (proportional to $F_{\mu\nu}^{(e)}$) one obtains the modification on the propagation of the photon mode transverse to the external magnetic field, whereas expanding the second part of this term (proportional to $\widetilde{F}_{\mu\nu}^{(e)}$) one gets the modification on the propagation of the mode parallel to the external field.  
So, the Euler--Heisenberg term introduces different refractive indexes for the transverse and parallel photon polarizations, giving rise to birefringence effects for the photon when it propagates in flat space.
For the cosmological case it should be noted that the Euler--Heisenberg Lagrangian breaks the conformal invariance of the electromagnetic field. So the modification of the photon wave equation due to the cosmic expansion will be of order
\begin{eqnarray}
\rho=\frac{4\alpha^2}{45m_e^4}\,.
\end{eqnarray}
Taking into account the results of the flat case \cite{Dolgov:2012be} and noting that the scaling of the magnetic field with the scale factor is such that $B^2=B_T^2/a^4$ \cite{Durrer:2013pga}, one should now have evolving refractive indexes of the form

\begin{eqnarray}
    n_{\perp}^2-1&=&4\rho B_T^2a^{-4}(\eta),\label{eq10}\\
    n_{\parallel}^2-1&=&7\rho B_T^2a^{-4}(\eta),\label{eq11}
\end{eqnarray}
where $ B_T $ is the part of the magnetic field transverse to the direction of the propagation of the wave. 
These terms can be encapsulated in an effective mass term in the wave equation. 
In principle, the above equations need to incorporate terms proportional to the Hubble rate. We neglect such contributions since we are only focusing into the leading effect of the Euler--Heisenberg 
term into the refractive indexes. This will not be the case for the analogous discussion about the gravitational refractive index as we will discuss.
Following reference \cite{Dolgov:2012be}, this effective mass is defined as

\begin{equation}
m_{\gamma}^2=\omega^2-k^2=\omega^2(1-n_{\gamma}^2).
\end{equation}
According to reference \cite{Dolgov:2012be} the effective mass in Minkowsky background, $\widetilde{m}_{\gamma}$, should be related to the effective mass in a cosmological background through $\widetilde{m}_{\gamma}=m_{\gamma}a^4(\eta)$. This agrees with the $a^{-4}$ factor that we have written in equations (\ref{eq10}) and (\ref{eq11}).
Taking into account this discussion, equation (\ref{eq7}) can be re-writen as
\begin{equation}\label{eq9}
    \left[\Box -m_{\gamma}^2\right]A^j=\kappa\partial_{i}\left(  h^{ik}F^{(e)j}_k-h^{jk}F^{(e)i}_k  \right).
\end{equation}
Equations (\ref{eq6}) and (\ref{eq9}) describe the graviton-photon system.

Before getting the final equations some considerations are in order. We redefine the gravitational wave tensor as $H_{ij}=ah_{ij}$. This definition allows us to remove the absorption term from equation (\ref{eq6}). We assume that GWs and EMWs expand in their Fourier modes such that
\begin{eqnarray}
    H_{ij}(z,\eta)&=&\sum_{\lambda}H_{\lambda}(\eta)e^{\lambda}_{ij}e^{ikz},\label{eq12}\\
    A_{j}(z,\eta)&=&i\sum_{\lambda}A_{\lambda}(\eta)e^{\lambda}_{j}e^{ikz},\label{eq13}
\end{eqnarray}
with $\lambda={\times,+}$ for the GW tensor and  $\lambda={\perp,\parallel}$ for the EMW vector. $\perp,\parallel$ denote modes perpendicular and parallel to the external magnetic field, respectively.  Assuming that the waves propagate in the direction of the z-axis, the non-zero elements of the tensor modes are defined such that $e^{+}_{xx}=-e^{+}_{yy}=1$ and $e^{\times}_{xy}=e^{\times}_{yx}=1$. Introducing the definitions (\ref{eq12}) and (\ref{eq13}) in equations (\ref{eq6}) and (\ref{eq9}), we obtain
\begin{eqnarray}\label{eq12}
\left[ \partial_{\eta}^2+\left( k^2-\dfrac{a^{''}}{a}\right) \right]H_{\lambda}=-\kappa akA_{\lambda}B_{T},\\
        \left(\partial_{\eta}^2+k^2  \right)A_{\lambda}-\left(\omega^2-k^2  \right)A_{\lambda}=-\dfrac{\kappa}{a}kH_{\lambda}B_T.\label{eq13}
\end{eqnarray}
Assuming that the Euler--Heisenberg term introduces small corrections, then the refractive index is close to unity, implying $|n_\gamma-1|\ll 1$; so, $\omega\approx k$. Hence, we can approximate the d'Alembert operator as, $(\partial_{\eta}^2+k^2)A_{\lambda}=(i\partial_{\eta}+k)(-i\partial_{\eta}+k)A_{\lambda}\simeq 2\omega(\omega+i\partial_{\eta})A_{\lambda}$, where we have also assumed $-i\partial_{\eta}A_{\lambda}\simeq \omega A_{\lambda}$.  In addition the effective mass term can be approximated as $(\omega^2-k^2)\simeq 2\omega^2(1-n_\gamma)$. On the other hand, for GWs we consider only the sub-Hubble modes, those that satisfy, $k\eta \gg 1$, so we can eliminate the term that goes with the second derivative in the scale factor of equation (\ref{eq12}). We also assume that the propagation of GWs do not differ much from the prediction in GR, that is $-i\partial_{\eta}H_{\lambda}\simeq \omega H_{\lambda}$.
With these approximations we get
\begin{eqnarray}\label{eq14}
        (\omega+i\partial_{\eta})H_{\lambda}=-\dfrac{\kappa}{2} aA_{\lambda}B_T,\\
    (\omega+i\partial_{\eta})A_{\lambda}+\omega(n-1)A_{\lambda}=-\dfrac{\kappa}{2}\dfrac{H_{\lambda}}{a}B_{T}.\label{eq15}
\end{eqnarray}
It can be noted that the mixing of the modes in this system, formed by equation (\ref{eq14}) and (\ref{eq15}), is not symmetric. However, if we consider again the physical mode $h(\lambda)$, we can obtain symmetric mixing. So, in order to obtain a system that is coupled by a symmetric matrix, we have to consider $A_{\lambda}$ and $h_{\lambda}$ as the degrees of freedom. We get
\begin{eqnarray}\label{eq16}
   (\omega+i\partial_{\eta}+i\mathcal{H})h_{\lambda}=-\dfrac{\kappa}{2}A_{\lambda}B_T,\\
    (\omega+i\partial_{\eta})A_{\lambda}+\omega(n-1)A_{\lambda}=-\dfrac{\kappa}{2}h_{\lambda}B_{T}.\label{eq17}
\end{eqnarray}
GWs and EMWs are coupled by these equations. There is an imaginary term in equation (\ref{eq16}) proportional to the Hubble parameter, which affects the propagation of gravitational modes. It acts as an effective gravitational refractive index term that causes absorption or amplification in the amplitude of the wave.

Now, regrouping the modes $A_{\lambda}$ and $h_{\lambda}$ in the state vector $\psi^{T}=\left[ A_{\lambda},h_{\lambda}\right]$, we can write the equations in matrix form and, therefore, identify the mixing matrix
  \begin{equation}\label{eq18}
      \left[  (\omega+i\partial_{\eta})+  \left(  \begin{matrix}
  \omega(n^{\lambda}_{\gamma}-1)  &
   \dfrac{\kappa B_T}{2}  \\
   \dfrac{\kappa B_T}{2} &
    \omega(n_{g }-1)
   \end{matrix}\right)  \right] \begin{bmatrix}
    A_{\lambda} \\
    h_{\lambda} 
  \end{bmatrix}=0,
  \end{equation}
where we define the effective refractive index of the gravitational wave as 
  \begin{equation}
n_g=1+i\dfrac{\mathcal{H}}{\omega},
  \end{equation} 
being the origin of this effective refractive index the expansion of the cosmological background.   It should be noted that equation (\ref{eq18}) reduces to that obtained in reference \cite{Dolgov:2012be} for propagation in a Minkowski background, that is $a=1$. In addition, it should be noted that in the case that the corrections introduced by the Euler--Heisenberg term are negligible, then $n_{\gamma}=1$ and, therefore, we recover graviton-photon conversion.


\subsection{Equations of motion beyond GR}\label{secIIB}
Let us now consider that the gravitational phenomena are described by an ATG. At this point we do not restrict attention to a particular theory. So, we consider a general parametrization for the linear TT perturbations for the tensor modes\footnote{This parametrization does not take into account theories that produce oscillations of gravitons with additional gravitational degrees of freedom present in the theory, as those treated in reference \cite{Jimenez:2019lrk}. Lorentz violating theories will be consider later in this section.}. This is \cite{Saltas:2014dha}
\begin{equation}\label{eq20}
    h^{''}_{ij}+2\mathcal{H}(1+\nu)h^{'}_{\ij}+(c_T^2k^2+\mu^2)h_{ij}=0.
\end{equation}
The parameter $\mu$ is the effective mass of the graviton, $c_T$ is the speed of GWs, and $\nu$ is a ``friction'' term responsible for variations in the amplitude of the tensor modes.  The posible presence of these parameters in equation (\ref{eq20}) will depend on the parameters of the particular theory. In some literature, the damping factor is called the running of the Planck mass, $\nu=\alpha_M=\mathcal{H}^{-1}d\log M_*^2/dt$, with $M_*$ the effective Planck mass. This term appears, for example, in theories with an extra field that is not minimally coupled to gravity. On the other hand, it has to be emphasized that according to reference \cite{LIGOScientific:2017vwq} the most accurate measurements of the speed of gravitational wave have an approximate value $c_T\simeq 1$ for the present Universe. So, we will take $c_T=1$.

As discussed in reference \cite{Cembranos:2018jxo} for the Minkowski case, when we compare equation (\ref{eq20}) with the corresponding equation of the graviton-photon system in GR, that is equation (\ref{eq6}), we can argue that in an ATG the graviton-photon system is also described by equation (\ref{eq18}) with the gravitational refractive index of the graviton given by
\begin{equation}
    n_g^2=1+i\dfrac{2\mathcal{H}(1+\nu)}{\omega}-\dfrac{\mu^2}{\omega^2}.
\end{equation}
Furthermore, we can also add a term $Ak^{\alpha}$ to this expression, taking into account a gravitational refractive index describing terms that modify the dispersion relation connected to Lorentz violations \cite{Mirshekari:2011yq}. Now, focusing again our attention to the case in which the propagation of GWs does not differ much from that predicted by GR, that is $n_g\approx 1$, we can write
\begin{equation}\label{eq22}
    n_g=1+i\dfrac{\mathcal{H}(1+\nu)}{\omega}-\dfrac{\mu^2}{2\omega^2}-\dfrac{A}{2}\omega^{\alpha-2}.
\end{equation}
The particular ATG will fix the value of the different terms appearing in this expression. 
In order to get a graviton-photon evolution equation of the form of equation (\ref{eq18}), we shall assume that the parameters $\nu$ and $\mu$ vary slowly or are constant in the interval under study. Moreover, as discussed in reference \cite{Cembranos:2018jxo}, when an ATG modifies the effective Plank mass (and, therefore, induces an attenuation parameter $\nu$) it changes the coupling with the photon modes in the wave equation of the graviton; so, we need to redefine the gravitational modes 
\begin{equation}\label{eq23}
\widetilde{h}_{\lambda}=\dfrac{\kappa}{\kappa_{eff}}h_{\lambda},
\end{equation}
with $\kappa_{eff}=M_{*}^{-1}$, to obtain a symmetric mixing matrix. Thus, we finally get
\begin{equation}\label{eq24}
      \left[  (\omega+i\partial_{\eta})+ \left(\begin{matrix}
  \omega(n^{\lambda}_{\gamma}-1)  &
   \dfrac{\kappa_{eff} B_T}{2}  \\
   \dfrac{\kappa_{eff} B_T}{2} &
    \omega(n_{g }-1)
   \end{matrix}\right)  \right] \begin{bmatrix}
    A_{\lambda} \\
   \widetilde{h}_{\lambda}
  \end{bmatrix}=0,
  \end{equation}
where the refraction index $n_g$ can be described by equation (\ref{eq22}). From now on we will omit the tilde in the gravitational mode. 

\subsection{Conversion probability in GR and beyond}\label{secIIC}
Equation (\ref{eq24}) describes the coupled evolution of the photon and graviton modes in the presence of an external magnetic field. The propagation of the modes depends on the refractive indexes that appear in the mixing matrix. We have shown that different terms, which are due to the cosmic expansion and/or the modification of the predictions of GR, can be included in the refractive index of the graviton. 
As the different polarizations do not mix with each other, the system can be reduced from four to two dimensions. Thus, one has
\begin{equation}\label{eq25}
\left[(\omega+i\partial_{\eta})+\mathcal M\right]\psi=0,
\end{equation}
with
\begin{eqnarray}\label{eq26}
\mathcal M=
\left({\begin{array}{cc}
   \Delta_ {\gamma} & \Delta_M  \\
   \Delta_M & \Delta_{\rm g} \\
  \end{array} } \right),\quad
\psi=\left(\begin{array}{c}
A_{\lambda}\\  h_{\lambda}
\end{array}\right),
\end{eqnarray} 
where the elements of the matrix are defined as
\begin{eqnarray}\label{eq27}
\Delta_M&=&B_T/M_{*},\nonumber \\
\Delta_{\gamma}&=&\omega(n_{\gamma}-1), \,\,  \Delta_g=\omega(n_g-1).
\end{eqnarray}
The photon refractive indexes are defined in equations (\ref{eq10}) and (\ref{eq11}), whereas the gravitational refractive index is given by equation (\ref{eq22}). Now, we decompose the gravitational refractive index in its real and imaginary parts, that is $\Delta_g=\Delta_g^R+i\Delta_g^I$. In order to solve the system we have to diagonalize the mixing matrix $\mathcal M$. As it is a symmetric matrix, it can be diagonalized as $\mathcal M'=O^T\mathcal M O$, $\Psi'=O^T\Psi$, being $O$ a complex
orthogonal matrix or complex rotation.  So, the elements for the $\mathcal M'$ matrix, which are the eigenvalues of 
$\mathcal M$, and the angle of rotation will be complex in general. These are
\begin{eqnarray}
m_{1,2}&=&\frac{1}{2}\left[\Delta_{\rm g}+\Delta_{\gamma}\pm\sqrt{4\Delta_{\rm M}^2+\left(\Delta_{\rm g}-\Delta_{\gamma}\right)^2}\right],\qquad\\
\tan(2\theta)&=&\dfrac{2\Delta_{\rm M}}{\Delta_{\rm g}-\Delta_{\gamma}}.
\end{eqnarray}
In this section we assume that $\mathcal M$ is approximately constant in the interval under consideration, in order to do an order of magnitude approximation. Solving the matrix equation we get
\begin{equation}
\begin{split}
A(\eta)=&\left[\cos^2\theta e^{im_1\eta}+\sin^2\theta e^{im_2\eta}\right] A(0) \\&-\sin\theta\cos\theta\left[e^{im_1\eta}-e^{im_2\eta}\right] h(0),\\
\end{split}
\end{equation}
\begin{equation}
\begin{split}
 h(\eta)=&\sin\theta\cos\theta\left[e^{im_1\eta}-e^{im_2\eta}\right]A(0)\\
&\left[\sin^2\theta e^{im_1\eta}+\cos^2\theta e^{im_2\eta}\right] h(0),
\end{split}
\end{equation}
where we have abserbed a phase $ e^{i\eta\omega}$ in the definition of the fields. Interpreting the perturbation of the fields as wave functions, the probability of finding a photon after a time $\eta$, $P_{{\rm g}\rightarrow{\gamma}}(\eta)=|A(\eta)|^2 =A(\eta)A^*(\eta)$, can be calculated for the case in which the initial state is formed only by gravitons. Defining the parameters
\begin{eqnarray}
\alpha=\dfrac{1}{2}\operatorname{Re}\left(\sqrt{4\Delta_{\rm M}^2+\left(\Delta_{\rm g}-\Delta_{\gamma}\right)^2}\right)\\
\beta=\dfrac{1}{2}\operatorname{Im}\left(\sqrt{4\Delta_{\rm M}^2+\left(\Delta_{\rm g}-\Delta_{\gamma}\right)^2}\right),
\end{eqnarray}
 the expression for the conversion probability takes the form
\begin{equation}\label{eq34}
P_{{\rm g}\rightarrow{\gamma}}(\eta)=
\dfrac{ \Delta_M^2 e^{-\Delta_g^I\eta}   }{\alpha^2+\beta^2}
\left[\sinh^2\left(\beta\eta\right)+\sin^2\left(\alpha\eta\right)\right].
\end{equation}
Taking $\Delta_g=0$ in this expression, we recover the predictions for GR in a Minkowski background \cite{Dolgov:2012be}. In that case resonance is reached for $\Delta_{\gamma}=0$ and the regime of weak mixing corresponds to $\Delta_{\gamma}\gg\Delta_{M} $. 
Even in GR we will have modifications of the vacuum probability for cosmic backgrounds, since in this case one has $\Delta_g=i\mathcal H$. So, for an expanding universe the probability will be damped. 
On the other hand, regarding ATGs, equation (\ref{eq34}) reduces to the conversion probability obtained in reference \cite{Cembranos:2018jxo} for a real gravitational refractive index, that is $\Delta_g^I=0$ and $\beta=0$. In this case resonance is achieved for a value of the mixing angle $\theta=\pi/4$, which implies $\Delta_{\gamma}=\Delta_g^R$, and $\Delta_g^I=0$; in a cosmological background this is only possible if $\nu = -1$, compensating the effect of the cosmic expansion.

We emphasize that the exponential factor of the conversion probability depends only on the attenuation part of $n_g$. For sufficiently low negative values of $\Delta_g^I$ the probability may grow up beyond unity. However, it has to be noted that this probability is normalized to the amplitude of the initial GWs and, therefore, the decoherence due to the interaction with the possible extra degrees of freedom of the ATG (or with the contraction of the universe if it were the case) can increase the probability to values larger than one. This effect can be better described with the density matrix formalism used in section \ref{secIII}.

The coupling term of the mixing matrix (\ref{eq26}) is determined by the present value of the cosmic magnetic field. In order to get an estimation, we assume as a first approximation that $M_*\simeq (16 \pi G)^{-1/2}$ in equation (\ref{eq27}). Then, we express the coupling term as in reference \cite{Dolgov:2012be}
\begin{equation}\label{eq35}
    \Delta_M =2.4\cdot10^{-15} \left[ \dfrac{B_T}{1G}\right]\qquad s^{-1}.\end{equation}
In reference \cite{Barrow:1997mj}, with measurements from the COBE experiment, a bound on the present value of the large-scale magnetic fields is found to be $B_T < 5\cdot10^{-9}$ G.
On the other hand, for the photon term we have written the refractive indexes for the different polarizations in equations (\ref{eq10}) and (\ref{eq11}). However, to have a correct description of the propagation of the photon one must consider an effective mass that arises from the effects of the electron primordial plasma on the photon, which depends on the plasma frequency as $ m_ {\rm plasma} = -\frac{\omega_{\rm plasma}^2}{2\omega}$. This effective mass is added to the QED contribution, such that, with the approximation made, $(n^2-1)\simeq 2(n-1)$, and for the values of the electromagnetic constants, $\Delta_{\gamma}$ can be expressed as \cite{Dolgov:2012be}
\begin{equation}\label{eq36}
\begin{split}
    \Delta_{\gamma}=&\Big| 4\cdot10^{-17}\left[ \dfrac{\omega}{1 eV}\right]\left[ \dfrac{B_T}{1G}\right]^2a^{-4}-
    \\
    &
    -1.05\cdot10^{-6}\left[ \dfrac{1 eV}{\omega}\right]\left[ \dfrac{n_e}{cm^{-3}}\right]^2\Big| \qquad s^{-1},
\end{split}
\end{equation}
where $n_e$ is the electron number density. 
 This expression for $\Delta_\gamma$ allows us to obtain an estimation for this term. Nevertheless, it should be noted that in order to get a more rigorous expression one should obtain the plasma effects on the photon field from a Lagrangian in a consistent way (along the lines, for example, of reference \cite{Ejlli:2016avx}).

We can compute the probability today considering the present values of the cosmological parameters. We take $n_e=2.47\cdot10^{-7} {\rm cm}^{-3} $ for the present electron density. For the current value of the magnetic field we use the upper bound  $B_0=5\cdot10^{-9}$ G. We consider primordial GWs propagating from the beginning of dark energy domination until the present time; this implies a total propagation time $\eta_0\simeq 3.7\cdot10^{17} $ s, during which we will assume the conformal Hubble rate today $\mathcal{H}\simeq 2.2\cdot10^{-18}  {\rm s}^{-1}$. So, the probability calculated for GR from (\ref{eq34}) is
\begin{equation}\label{eq37}
    P_{{\rm g}\rightarrow{\gamma}}^{\rm GR}(\eta_0)=8.6\cdot10^{-12},
\end{equation}
where we have taken $\Delta_g=i\mathcal{H}$  for the graviton.  For the frequency spectrum we have chosen the largest frequency within the valid range of the Euler--Heisenberg approximation, this is $\omega=10^5$ eV, $w <m_e$ with $m_e=5\cdot10^5$ eV. This is the frequency for which the probability has the maximal value. 
As it can be seen in figure \ref{fig1}, the mixing probability oscillates and increases with frequency.

We have to emphasize that this probability has been calculated keeping the parameters of the mixing matrix constant during the evolution, as a first approximation. This rude approximation allows us to understand the dependence of the probability with modifications of the predictions of GR, as we will discuss now for some examples.
 
 \subsubsection{Massive Gravity}
Let us consider a theory of massive gravity. In this case the gravitational refractive index has an additional real term apart from the imaginary part coming from the cosmic evolution. 
So, the gravitational term appearing in the mixing matrix can be written as
\begin{equation}\label{eq38}
    \Delta_g=-2.4\cdot10^{14}\left[\dfrac{\mu}{1\, {\rm eV}}\right]^2\left[\dfrac{1\,{\rm eV}}{\omega}\right]+i\left[\dfrac{\mathcal{H}}{1\,{\rm s}^{-1}}\right] \quad {\rm s}^{-1}.
\end{equation}
Nowadays there are several bounds on the mass of the graviton coming from different observations. Supernova data put the limit in $\mu <10^{-23}$ eV and LIGO GWs measurements lower it to $\mu<10^{-29}$ eV \cite{Will:2014kxa}. It has to be noted that these bounds do not take into account the new kinds of polarization that arise for a non-vanishing mass for the graviton. In massive gravity there are (at least) 5 polarizations that can have effects on gravitational tests in the Solar System or in gravitational lensing, although they cannot be detected by current interferometers. As stated in reference \cite{deRham:2016nuf}, the precession of the perihelion of the lunar orbit establishes $\mu<10^{-30}$ eV. Other measurements of the Earth-Moon system establish the bound in $\mu<10^{-32}$ eV and for the moment the lowest limit taken by gravitational lensing is $\mu<10^{-33}$ eV. 

Assuming a theory of massive gravity that saturates the lowest bound, that is $\mu=10^{-33}$ eV, and taking the present values for the parameters already discussed in the GR case, equation (\ref{eq34}) leads to
\begin{equation}\label{eq38}
P_{{\rm g}\rightarrow{\gamma}}(\eta_0)=8.6\cdot10^{-12},
\end{equation} 
which is the same value as the one obtained for GR, that is equation (\ref{eq37}).  Therefore, a non-vanishing graviton mass with a value within the limits established by current observations does not have a direct effect in the graviton-photon mixing probability.

\subsubsection{Theories modifying the attenuation term}
A simple example of theories introducing an attenuation term for the propagation of GWs are $f (R)$ theories of gravity. This term introduces an imaginary part for the index of refraction (\ref{eq22}). Let us assume that there is neither mass term for the graviton nor Lorentz violation, so the gravitational index of refraction is purely imaginary. We use again the values presented in equations (\ref{eq35}) and (\ref{eq36}) and write the graviton term as
\begin{equation}\label{eq40}
    \Delta_g=i(1+\nu)\mathcal{H}\qquad s^{-1},
\end{equation}
which is independent of $\omega$. In reference \cite{Belgacem:2018lbp} the authors consider a non-local theory of gravity that only affects the propagation of GWs in a $\nu$  term, as a typical $f(R)$ theory. The value of $\nu$ calculated from measurements of the Hubble parameter of GW170817 in LIGO is  $\nu=5.1_{-20}^{+11}$  \cite{Belgacem:2018lbp} . For illustrative purposes, we can compute the values of the probability implied by a theory that saturates the upper observational bound and another saturating the lower bound. Those are 
\begin{equation}
  P_{{\rm g}\rightarrow{\gamma}}(\eta_0)=2.09\cdot10^{-13},
\end{equation}
for the upper bound and,
\begin{equation}
         P_{{\rm g}\rightarrow{\gamma}}(\eta_0)=1.67,
\end{equation}
for the lower one. As we can see, the values differ 13 orders of magnitude. We have taken again the frequency $\omega=10^5$ eV. 

\begin{figure}[!htb]
\begin{center}
\includegraphics[width=3in]{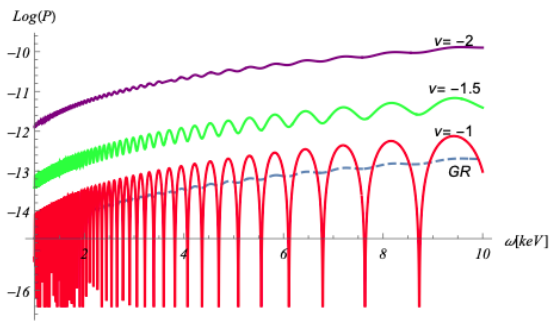}
\caption{The probability in logarithmic scale is represented as a function of the frequency, in the range of $0-10$ keV. We compute it for different values of $\nu$ and we also include the case of GR ($\nu=0$). The values of the parameters are assumed constant as $B_0 =5\cdot10^{-9}$ B, $n_e=2.47\cdot10^{-7} {\rm cm}^{-3} $ and $\mathcal{H}=2.2\cdot10^{-18}  {\rm s}^{-1}$.}
\label{fig1} 
\end{center}
\end{figure}

In figure \ref{fig1} we show the probability in terms of the frequency spectrum for different values of $\nu$. The oscillation is stronger for values closer to $\nu=-1$, reaching zero conversion. For negative values of $\nu$ with larger absolute values, the amplitude of the probability increases and the oscillation is suppressed. Note that $\Delta_g/\omega\ll 1$ for $\nu_{max}$ and, therefore, we are in the range of the Euler--Heissenberg approximation.

It should be noted that the values for the parameter $ \nu $ that we have used should be taken just as an approximation. Those values are obtained with a very low redshift, with a single source, and the range of the measurement is too long for the value obtained to be useful except to study its behavior with $\nu$. The probability is quite sensitive to $\nu$, reaching very high quantities for very large negative values. It must be emphasized that the probability is not normalized and can rise above unity. Furthermore, the cases of an imaginary $n_g$ are more properly studied with the density matrix formalism, as we cannot approximate the system as being closed (that is, independent of the interaction with other degrees of freedom or even with the cosmic evolution). So, the dependence of the probability with $\nu$  will be better understood with the density matrix formalism. More importantly, with that formalism we will not approximate $\mathcal{H}$ as a constant parameter.



\section{Density matrix formalism}\label{secIII}
In the previous section we have treated the graviton-photon system as a closed system.
However, we know that in the primitive universe the medium was composed of a dense plasma that can interact with EMWs, reducing their interaction with GWs in our system.       
In the recombination epoch, at redshift $z\simeq1090$, the collisions of photons with free electrons were decreasing, but some effects must still be considered. These collisions would produce decoherence effects and the system would become an open system. 
When the length of oscillation is less than the average free path of the photons, it can be neglected; nevertheless, it has to be taken into account while the expansion of the universe keeps them interacting.
In order to introduce the decoherence effects in the system, we will apply the density matrix formalism in this section. In this framework, the decoherence can be introduced by inserting an antihermitic matrix into the Hamiltonian \cite{Dolgov:2012be}. 

On the other hand, it has to be noted that GWs can interact with additional fields. In particular, we have shown that the cosmic expansion and the additional degrees of freedom of some ATGs introduce a complex term in $n_g$. This term give rise to attenuation effects and can be interpreted as a signal of decoherence analogous to a ``plasma" for gravitational radiation. So, gravitational decoherence should also be taken into account in the density matrix formalism.

 \subsection{Formalism}\label{secIIIA}
 The evolution of the density matrix is derived from the Schr\"odinger equation and is given by the Liouville-von Neumann equation
 \begin{equation}
     i\dfrac{d\rho}{dt}=\left[  \hat H,\rho  \right]\,.
 \end{equation}
  We take equation (\ref{eq25}) as our Schr\"odinger equation and define the density matrix as $\rho=\psi\psi^{\dagger}$, with $\psi^{T}=\left[ A_{\lambda},h_{\lambda}\right]$. The Hamiltonian of the system is given by the mixing matrix $\mathcal M$. We introduce the photon decoherence effects into the system defining the hermitian matrix, $ \Gamma $, such that the new Hamiltonian takes the form
  \begin{equation}
     H =\mathcal M-i\dfrac{1}{2}\Gamma,
  \end{equation}
  where $ \Gamma $ is given by the collision integrals of the Boltzmann equation. It only interacts with the electromagnetic part; so, at first linear order, the matrix is written as
    \begin{equation}
    \Gamma=\left({\begin{array}{cc}
   \Gamma_{\gamma} &0 \\
   0 & 0 \\
  \end{array} } \right),
  \end{equation} 
being $\Gamma_{\gamma}=\sigma_Tn_e$ the inverse of the mean free path of photons due to Thompson scattering with electrons and $\sigma_T $ the Thompson cross section. Decoherence is represented as an antihermitic part of the Hamiltonian, so we have to take into account that the mixing matrix $\mathcal M$ already has an antihermitic part contained in the complex parameter $ \Delta_g $. Then, taking it into account, from equation (\ref{eq24}), we write the evolution of the density matrix as
  \begin{equation}\label{eq48n}
           i\dfrac{d\rho}{d\eta}=\left[  \mathcal M_{(+)},\rho  \right]+\{  \mathcal M_{(-)}-\dfrac{i}{2}\Gamma,\rho  \},
  \end{equation}
  where $ \mathcal M_{(+)}$ and $\mathcal M_{(-)}$ are the hermitic and antihermitic part of $\mathcal M$, respectively, $\mathcal M_{(-)}$ being given by the complex part of $n_g$.
 As in the case of $\Gamma$ for EMWs, the antihermitic part of $\mathcal M$ is equivalent to a damping factor for GWs due to the interaction with external degrees of freedom.

In both matrices, $\mathcal M$ and $\Gamma$, there are parameters that depend on the scale factor. While in the wave function formalism we have calculated the probability by treating them as approximately constant in a short time interval, in this formalism we will write their dependence explicitly and calculate the evolution in terms of the scale factor. With this aim, we first express equation (\ref{eq48n}) using the scale factor $a$ instead of the conformal time $\eta$; so that the evolution of $\rho$ is now given by
\begin{equation}\label{eq47}
 i\mathcal{H}a\dfrac{d\rho}{da}=\left[  \mathcal M_{(+)},\rho  \right]+\{\mathcal M_{(-)}-\dfrac{i}{2}\Gamma,\rho  \}.
\end{equation}
We describe the elements outside the diagonal in terms of their real and imaginary parts, so $\rho_{\gamma g}=\rho_{g \gamma}^*=R+iI$,  while the terms in the diagonal, $\rho_{\gamma}$ and $\rho_{g}$, are real numbers.   The diagonal elements of $\rho$ represent the probabilities of finding a photon or a graviton, repectively. Then, the probability of graviton to photon conversion will be given by the value of $\rho_{\gamma}$ defining an initial state without photons. From equation (\ref{eq47}) we obtain the following system of differential equations
\begin{eqnarray}\label{eq48}
\dfrac{d\rho_{\gamma}}{da}&=&\dfrac{2\Delta_M I+2\Gamma_{\gamma}\rho_{\gamma}}{\mathcal{H}a},\\
\dfrac{d\rho_{g}}{da}&=&-\dfrac{2\Delta_M I+2\Delta_g^I \rho_g}{\mathcal{H}a},\label{eq49}\\
\dfrac{dR}{da}&=&\dfrac{(\Delta^R_{g}-\Delta_{\gamma})I-(\Delta_g^I+\Gamma_{\gamma}) R}{\mathcal{H}a},\label{eq50}\\
\dfrac{dI}{da}&=&\dfrac{(\rho_g-\rho_{\gamma})\Delta_M+(\Delta_{\gamma}-\Delta^R_{g})R-(\Delta_g^I+\Gamma_{\gamma}) I}{\mathcal{H}a}. \label{eq51}\qquad
\end{eqnarray}

Now we must consider the dependence in the scale factor of the parameters we have introduced in these equations, which are the damping factor and the plasma mass of the photon. These are
  \begin{eqnarray}\label{eq52}
    \Delta_{\gamma}(a)&=&3.36\cdot10^{-4}X_e(a)\left[ \dfrac{1 {\rm eV}}{\omega_i}\right]\left[ \dfrac{n_{ie}}{{\rm cm}^{-3}}\right]^2\left( \dfrac{a_i}{a}\right)^2, \qquad\\
    \Gamma_{\gamma}(a)&=&6.36\cdot10^{-12}X_e(a)\left( \dfrac{a_i}{a}\right)^3 \quad s^{-1},\label{eq53}
\end{eqnarray}
  where the subscript $i$ refers to the initial states at the time of recombination. We take as reference the scale factor $a_i = 1$. So the scale factor at present would be $ a_0 = 1090 $. We have introduced the fraction of electron ionization in the definition of electronic density, $n_e(a)=X_e(a)n_{Bi}(\frac{a_i}{a})^3$, with $n_B$ the baryonic density, which in recombination has a value $n_{Bi}=320 $  ${\rm cm}^{-3}$. On the other hand, the ionization fraction is predicted by the knowledge of atomic physics and cosmology. Its evolution is governed by a differential equation which is solved  numerically \cite{Dolgov:2012be}. 
\begin{figure}[h]
\begin{center}
\includegraphics[width=3in]{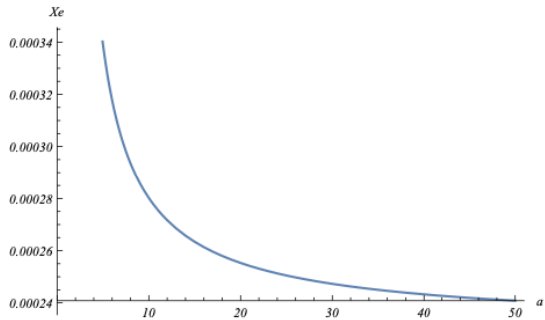}
\caption{Decrease in the fraction of free electron density with respect to the baryonic density as a function of the scale factor.  It is calculated numerically as in reference \cite{Dolgov:2012be}.}
\label{fig2} 
\end{center}
\end{figure}
In figure \ref{fig2} the fraction of free electron density with respect to the baryonic density is depicted for values up to the reionization period, which starts with the birth of the first stars at $z \sim 20$. At that moment the fraction of ionization increases rapidly up to $X_e \sim 1$, leaving the medium again ionized due to the radiation of the stars, and it remains that way until the present time.

On the other hand, since the mass of the graviton is limited to very low values and we have seen in the previous section that it does not have an important effect in graviton-photon mixing, we set $\Delta_g^R =0$. So, in this section we just focus on the attenuation part of the gravitational refractive index, neglecting also potential Lorentz violating effects.
Thus, $\Delta_g$ is given by equation (\ref{eq40}), that is
\begin{equation}\label{eq55}
    \Delta_g=i(1+\nu)\mathcal{H}\qquad s^{-1}.
\end{equation}
It should be noted that the term $\nu$ could depend on $a$ through the additional degrees of freedom of the ATG.

Finally, the Hubble parameter can be given assuming a background evolution compatible with a $\Lambda$CDM model\footnote{It should be noted that in ATG the background cosmic evolution could differ from the standard $\Lambda$CDM model. However, we assume that potential discrepancies are not observable at the background level.}. After recombination, we can express it as
  \begin{equation}\label{eq55}
      \mathcal{H}(a)=8\cdot10^{-14}\left( \dfrac{\Omega_M}{a}+7.7\cdot10^{-10}\Omega_{\Lambda}a^2 \right)^{1/2}\quad {\rm s}^{-1},
  \end{equation}
where we have normalized $a=1$ at recombination, whereas we have taken $\Omega_M=0.3$ and $\Omega_{\Lambda}=0.7$ as approximate values for the density parameters.

 \subsection{Phenomenology}\label{secIIIB}
We already have all the terms needed to solve the graviton-photon mixing system. We cannot obtain an analytic solution for $\rho_{\gamma}$ but we can solve system (\ref{eq48})-(\ref{eq51}) numerically taking into account equations (\ref{eq40}), (\ref{eq52}), (\ref{eq53}), and (\ref{eq55}) with values of phenomenological interest.

\begin{figure}[h]
\begin{center}
\includegraphics[width=3in]{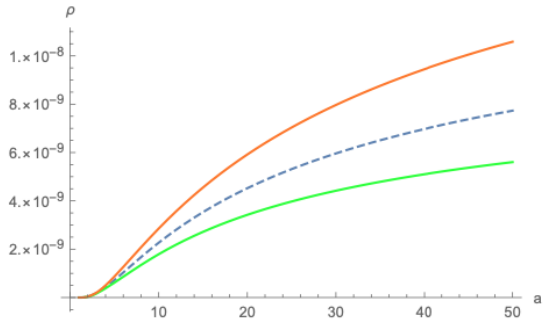}
\caption{Graviton-photon conversion probability as a function of the scale factor for a graviton with frequency $\omega=10^5$ eV and external magnetic field $B=3\cdot10^3$ G. We define the initial values $\rho_g(1)=1$ and $\rho_{\gamma}(1)=R(1)=I(1)=0$.  The period of evolution is from the initial state $a=1$ at recombination to $a=51,9$, when reionization begins. It has been calculated solving the system (\ref{eq48})-(\ref{eq51}) numerically. The dashed line correspond to the GR case with $\nu=0$, the green line corresponds to  $\nu=0.06$ and the orange line to $\nu=-0.06$.}
\label{fig3} 
\end{center}
\end{figure}

\begin{figure}[h]
\begin{center}
 \includegraphics[width=3in]{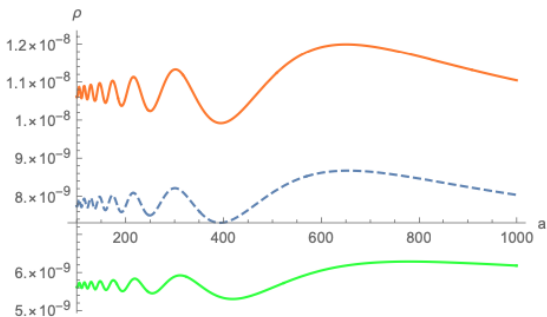}
\caption{This figure represents the continuation of figure 3 for the evolution of probability. Three cases are depicted: GR that is $\nu=0$ (dashed), $\nu=0.06$ (green), and $\nu=-0.06$ (orange). The period shown begins at reionization, $a = 51.9$, and ends at the present time, $ a = 1090 $.}
\label{fig4} 
\end{center}
\end{figure}

Let us first discuss the results for GR, that are obtained taking $ \nu = 0 $ in equation (\ref{eq40}). The evolution of the conversion probability in GR as a function of the cosmological scale factor can be seen in figures \ref{fig3} and \ref{fig4} (dashed line). Two phases can be clearly differentiated during the evolution of the conversion probability characterized by the shape of the ionization fraction $ X_e $.  The first period corresponds to the decrease in the amount of free electrons in the Universe due to the recombination with the protons, figure \ref{fig3}. During this phase the probability increases with time, until the period of reionization begins. From that moment, when the medium is again ionized, the probability begins to oscillate as it is shown in figure \ref{fig4}.  The oscillations have an amplified amplitude due to the imaginary gravitational refractive index characterizing cosmological backgrounds in GR. Its effect dominates over the damping of $\Gamma_{\gamma}$, which tends to attenuate the probability. Such amplification is not observed in Minkowski's spacetime \cite{Dolgov:2012be}. 


From the previous results for GR and the estimation in ATGs using the wave function formalism discussed in the previous section, we can deduce that the probability is sensitive to the amplification/attenuation term of the graviton.  In order to integrate the system for ATGs, we will consider that $\nu$ is constant or varies slowly in the interval of interest, so that it can be treated as a parameter characteristic of the ATG. The value of $\nu$ is bounded by observations  \cite{Belgacem:2018lbp}, as already discussed in the previous section. In figures \ref{fig3} and \ref{fig4} we represent the evolution of the probability for different values of $\nu$. Comparing with the GR case, we observe that the same phases in evolution are present.   For values of $\nu<0$ the probability is amplified, starting with a faster increase in the first phase and, in the second phase, an increase in the amplitude of oscillations, as opposed to what happens for values of $ \nu> 0 $.
As we have already discussed using the wave function formalism, the probability is highly sensitive to the value of $\nu$.

\section{Conclusion}\label{secIV}
We have studied the phenomenon of graviton-photon oscillation in a cosmological background in GR. 
%
We have argued that the prediction for the mixing probability of GWs after recombination (with energies of the order of $10^5$ eV at that time) is of order  $P_{g\longrightarrow\gamma}\sim10^{-11}$. This value is in agreement with that obtained in reference \cite{Dolgov:2012be}, even if we take into account the cosmic expansion.

We have also considered the possibility that an ATG could describe the gravitational phenomena. The new parameters introduced by the ATG can be taken into account in the graviton-photon system through an effective refractive index for the graviton.  
Applying the wave function formalism, and within the allowed range from different observations, we have concluded that a mass term for the graviton does not alter the general relativistic predictions for graviton-photon oscillation. On the other hand, we have shown that the mixing probability has a strong dependence on the attenuation term, which only affects the imaginary part of the gravitational refractive index.  We have compared the mixing probability for different values of the attenuation term $ \nu$ and concluded that it decreases for $\nu>0$ and increases for $\nu<0$, being the amplitude of the oscillations amplified for lower negative values of  $ \nu $. It can be noted that for $ \nu = -1 $ one has $ \Delta_g^I=0 $, as if decoherence be effectively removed. This particular case corresponds to resonance.

Furthermore, we have discussed the consequences of having an open graviton-photon system using the density matrix formalism. With the intuition acquired using the wave function formalism, we have focused on the most interesting case of ATGs leading to purely imaginary deviations of the GR gravitational refractive index. Assuming an approximately constant attenuation term, we have integrated the system numerically and confirmed the amplification of the effects for theories with $\nu<0$.

Although there is no doubt of the qualitative results presented in this work, a more rigorous treatment should take into account the effect of the cosmic expansion on the parameters of the model.
For example, a particular ATG would predict a concrete evolution for the attenuation term, which can be substituted when integrating the system for a particular model.
Whereas we have preferred to discuss graviton-photon mixing in a model independent way, setting out bounds corresponding to the allowed range of values for the parameters, that study may serve to establish new constraints on the parameters of different ATGs. 
Moreover, in this work we have not taken into account a general Lagrangian term for the photon interacting with other fields. In particular, we have not described how the plasma effects on the photon field come from a Lagrangian term. Those effects should be consistently described following the spirit of reference \cite{Ejlli:2016avx}.


\begin{acknowledgments}
This work was partially supported by the MICINN (Ministerio de Ciencia e Innovacion, Spain) project PID2019-107394GB-I00/AEI/10.13039/501100011033 (AEI/FEDER, UE) and the COST (European Cooperation in Science and Technology) Actions CosmicWISPers CA21106 and CosmoVerse CA2136. JARC acknowledges support by Institut Pascal at Universite Paris-Saclay during the Paris-Saclay Astroparticle Symposium 2022, with the support of the P2IO Laboratory of Excellence (program ``Investissements d'avenir" ANR-11-IDEX-0003-01 Paris-Saclay and ANR-10-LABX-0038), the P2I axis of the Graduate School of Physics of Universite Paris-Saclay, as well as IJCLab, CEA, APPEC, IAS, OSUPS, and the IN2P3 master projet UCMN.
\end{acknowledgments}

\end{document}